\documentstyle[prl,epsfig,aps]{revtex}
\begin{document}
\preprint{LPTENS-97/18}

\twocolumn[\hsize\textwidth\columnwidth\hsize\csname@twocolumnfalse\endcsname

\title{From second to first order transitions in  a disordered quantum magnet}
\author{L. F. Cugliandolo$^1$, D. R. Grempel$^{2,}$\cite{add1} and 
C. A. da Silva Santos$^3$ }
\address{ $^1$
\it Laboratoire de Physique Th{\'e}orique de l'{\'E}cole Normale 
Sup{\'e}rieure,
24 rue Lhomond, 75231 Paris Cedex 05, France and \\
 Laboratoire de Physique Th{\'e}orique  et Hautes {\'E}nergies, Jussieu, 
5{\`e}me {\'e}tage,  Tour 24, 4 Place Jussieu, 75005 Paris, France
\\
$^2$ CEA-Service de Physique de l'{\'E}tat Condens{\'e}, CEA-Saclay,
91191 Gif-sur-Yvette CEDEX, France
\\
$^3$
Laboratoire de Physique Th{\'e}orique et Mod{\`e}les Statistiques, B{\^a}t. 100,
Universit{\'e} Paris-Sud, Orsay, F-91405, France
}

\date\today
\maketitle
\begin{abstract}
We study the spin-glass transition in a disordered quantum 
model. There is a  region in 
the phase diagram where 
quantum effects are small and the phase transition is second
order, as in the classical case. In another region, quantum
fluctuations
 drive the transition first order. Across the first order line   
the susceptibility is discontinuous and shows hysteresis. Our findings
  reproduce qualitatively observations on  LiHo$_x$Y$_{1-x}$F$_4$.
We also discuss a marginally stable spin-glass state and derive some 
results previously obtained from the real-time dynamics of the model coupled to a bath.
\end{abstract}
\twocolumn 
\vskip .5pc]
\narrowtext

The study of quantum effects on the properties of spin glasses is a
subject of great experimental and theoretical interest. Spin-glass phases have been identified in 
systems such as mixed hydrogen-bonded 
ferroelectrics \cite{Pirc}, the dipolar magnet 
 LiHo$_x$Y$_{1-x}$F$_4$ \cite{aeppli1,aeppli2} or Sr-doped
La$_2$CuO$_4$ \cite{shirane} where quantum mechanics plays a
fundamental role. An important question is whether 
quantum spin glasses are 
 qualitatively different
from their classical 
counterparts  at low temperature. There is growing experimental evidence that the answer
to this question is affirmative  
both in and out of equilibrium \cite{aeppli1}. 
The thermodynamics of several models of disordered 
magnetic systems has been investigated with various techniques. 
Mean-field-like models have been solved  
using the replica formalism 
in imaginary-time
\cite{Brmo,Yeresa,kopec,daniel,oppermann,Gepasa,Niri,Sachdev,Shsi} and the Ising 
model in a transverse field has also been studied in finite dimensions\cite{Fisher,heiko,young}.
It is generally found that, in terms of a suitably defined quantum
parameter $\Gamma$, a boundary $\Gamma_{\rm c}(T)$ in the
$\Gamma-T$ plane separates spin-glass (SG) and paramagnetic (PM) phases. The
transition line ends at a quantum critical point at $T=0,\,\Gamma_{\rm
c}(0)$  above which the system is paramagnetic at all temperatures.
In the case of the quantum spherical $p$-spin model, the real-time
dynamics of the system coupled to a phonon bath  
was also investigated \cite{Culo}.
In this case, a boundary $\Gamma_{\rm d}(T)$ was found  across which
there is a {\it dynamic} phase transition from a PM 
state with equilibrium dynamics to a SG with 
non-stationary, aging, dynamics. 

In this paper we investigate in detail the equilibrium properties of 
this model. We find that a
tricritical point $(T^{\star},\Gamma^{\star})$ divides the line
$\Gamma_{\rm c}(T)$ in two parts. For $T \ge T^{\star}$, the SG
transition is of second order and the behavior of the 
quantum system is qualitatively similar to that of the classical one. However,  for $T < T^{\star}$ quantum
fluctuations drive the transition first order. The magnetic susceptibility is discontinuous and shows 
hysteresis across the first-order line. These findings reproduce
qualitatively the observed
behavior of LiHo$_x$Y$_{1-x}$F$_4$ in a transverse magnetic
field \cite{aeppli1,aeppli2}. The equations describing this system are non-linear and there is multiplicity of solutions in  
parts of the phase diagram. We found as a surprise that the usual
criteria used to choose between them have to be reinterpreted in order to get physically meaningful
solutions in the region $T < T^{\star}$. 
We also discuss the properties of solutions obtained through the use
of the {\it marginality condition}, an approach recently applied to the study 
of quantum 
problems\cite{Gepasa,Gile}. It is known from
work on  classical models \cite{marginality} that the
results of  
this approach are closely related to those obtained from the analysis of
the real-time dynamics of the  system. We explicitly show that this holds
true in our quantum case. This unables us to identify the dynamical
transition line $\Gamma_{\rm d}(T)$ and to derive 
certain properties of the non-equilibrium dynamics using the replica
calculation.

The  Hamiltonian of the quantum $p$-spin spherical model is 
\begin{equation}
H[{\bf P},{\bf s},J]= {1\over 2M} \sum_{i=1}^N P_i^2 - \sum^{N}_{i_1<...<i_p} J_{i_1...i_p}
s_{i_1} ... s_{i_p}
\; ,
\label{eq:action}
\end{equation}
where $s_i$ is a scalar spin variable and the conjugated momenta $P_i$
satisfy the commutation relations $[P_i,s_j] = -i \hbar \delta_{ij}$. 
A Lagrange  multiplier $z$ enforces the 
spherical constraint $1/ N\sum^{N}_{i=1} \langle s_i^2\rangle = 1$.  
The interactions $J_{i_1...i_p}$ are chosen from a Gaussian distribution 
with zero mean and variance
$
[J_{i_1...i_p}^2]_J = {\tilde{J}}^2 p!/(2N^{p-1})
$.
The model has  glassy properties for all $p\geq 2$. 
The Hamiltonian (\ref{eq:action}) may be interpreted in several ways. 
It represents a non-linear generalization of the quantum-rotor spin-glass models discussed in the literature
\cite{Sachdev}.  It also describes a quantum particle moving in an
$N$ (eventually infinite) dimensional space in the presence of a
random potential.
Finally, its partition function is formally identical to that of  
a classical chain of ``length'' $L=\beta\hbar$ embedded in an $N$-dimensional 
random environment.

The equilibrium properties of the model are obtained using a replicated imaginary-time 
path integral formalism \cite{Brmo}. In the large $N$ limit, the saddle-point 
evaluation of the partition-function allows us to define the order-parameter 
$Q_{ab}(\tau-\tau') = 1/N\sum_{i=1}^N \langle {\cal T} s_i^a(\tau)
s_i^b(\tau') \rangle$ where $a,b=1,\dots,n$ denote the replica indices and ${\cal T}$ 
the imaginary-time ordering operator. 
In terms of $Q_{ab}$ the free-energy per spin reads
\begin{eqnarray}
\nonumber
{\rm F}
&=&
\lim_{n\to 0} \frac1{2 n} 
\left\{ -\frac{1}{\beta} \sum_{\omega_k} 
\left[
\mbox{Tr} \ln \left( \frac{\tilde Q(\omega_k)}{\beta\hbar} \right) -
        n \left( (M \omega_k^2 + z) \right.\right.\right.
\\
&\times &\left. \left. \left.\frac{\tilde q_d(\omega_k)}{\hbar} -1 \right) 
\right]
- 
nz
-\frac{\tilde J^2}{2\hbar} 
\sum_{ab} \int_0^{\beta\hbar} d\tau  \; Q^p_{ab}(\tau) 
\right\}
\label{eq:free-energy}
\end{eqnarray}
\noindent
where 
$\beta=1/(k_B T)$ is the inverse temperature,  
$\omega_k=2\pi k/\beta$ are the Matsubara frequencies, 
$\tilde Q_{ab}(\omega_k)=\int_0^{\beta\hbar} d\tau
Q_{ab}(\tau)e^{i\omega_k \tau}$ 
and
$\tilde q_d(\omega_k)=\tilde Q_{aa}(\omega_k)$.
>From here on we take $\tilde J$ as the unit of energy, $\hbar/\tilde J$
as the unit of time, and work with dimensionless quantities. Quantum
fluctuations are controlled by the 
parameter $\Gamma\equiv \hbar^2/(\tilde J M)$. The classical limit of
the model\cite{Crso} is recovered when $\Gamma\to 0$. 
The equilibrium solutions are determined by requiring that 
$\tilde Q_{ab}(\omega_k)$, parametrized according to different {\it ansatze}, 
be an extremum of ${\rm F}$.  In the following we concentrate on the case $p \ge 3$. The phenomenology of the $p=2$ case \cite{Sachdev,Shsi} is not as rich.

For sufficiently high $T$ and/or $\Gamma$, 
thermal and/or quantum fluctuations destroy the SG phase and the system is in the PM phase.
The free-energy is then extremal for $Q_{ab}(\tau) =\delta_{ab}
q_d(\tau)$. Its Fourier transform is the solution of the equation
\begin{eqnarray}
\label{eq:para}
\tilde{q_d}(\omega_k) = \left[\omega_k^2/\Gamma + z
-\Sigma(\omega_k)\right]^{-1},
\end{eqnarray}
with $\Sigma(\omega_k) = p/2 \int_0^{\beta} d\tau 
q_d^{p-1}(\tau) e^{i\omega_k\tau}$ and $z$ is determined from 
$q_d(0)=1$. The above equation is non-linear and may have several
solutions, some of which may be spurious. 
We solved  Eq.\,(\ref{eq:para}) numerically for $p=3$.  We found that
for $T>T^{\star}\approx 1/6$ there is only one solution,
irrespective of the value of $\Gamma$. However, for $T<T^{\star}$, {\it three} solutions coexist in a finite
region of the $T-\Gamma$ plane (not including the $\Gamma$=0
axis). One of them 
is unstable and can be discarded from the start. 
We discuss below how to choose the physical solution between
the remaining two.
In the SG phase, inspired by the classical case \cite{Crso}, we searched for one-step RSB solutions of the form $Q_{ab}(\tau)=q'_d(\tau)
\delta_{ab} + q_{\rm EA}\epsilon_{ab}$, 
where $\epsilon_{ab}=1$ if $a$ and $b$ belong to the same diagonal block of size $m\times m$ and 
zero otherwise, and  $q'_d(\tau) = q_d(\tau) - q_{\rm EA}$.
The diagonal part, $q_d(\tau)$, the breaking point, $m$, and the
Edwards-Anderson order parameter, $q_{\rm EA}$, are determined by 
extremizing ${\rm F}$.
We find 
\begin{equation}
\label{eq:sg}
\tilde{q}'_d(\omega_k) = \left[\omega_k^2/\Gamma +  z' -
(\Sigma'(\omega_k)- \Sigma'(0))\right]^{-1},
\end{equation}
where $\Sigma'(\tau)= p/2 (q^{p-1}(\tau)-q_{\rm EA}^{p-1})$, $z' =p/2 \beta m q_{\rm EA}^{p-1}
(1+x_p)/x_p$ and 
\begin{equation}
\label{eq:theta}
m =T x_p\,  \sqrt{2/(p (1+x_p))} q_{\rm EA}^{-p/2}.
\end{equation}
The parameter $x_p$, solution of an algebraic equation, depends on $p$
only, and we found $x_3=1.817$. The condition $q_d(0)=1$ now yields an equation for the 
breaking point of the form   
$m\equiv \mu _T(\Gamma)$. Solutions of Eq.\,(\ref{eq:sg}) exist only for
$\Gamma \le \Gamma_{\rm max}(T)$. Above this value, quantum
fluctuations destroy the SG phase. We found that the function $\mu _T(\Gamma)$ has two real branches
in the interval $0\le \Gamma \le \Gamma_{\rm max}(T)$. The physical values
of $m$ are 
on the lowest branch which verifies $\mu _T(0)=m_{\rm class}(T)$, the
classical 
breaking point parameter. The situations  above and below
$T^{\star}$ are different. For $T \ge T^{\star}$ (but
lower than the classical transition temperature), $m_{\rm max}\equiv m(\Gamma_{\rm
max}) = 1$, its largest possible value. For $T < T^{\star}$, instead,  $m_{\rm
max} <  1$. In both cases, $q_{\rm EA}$ is finite at $\Gamma_{\rm
max}(T)$. We discuss below the consequences of these facts. We found
that $\lim_{T\to 0}  m_{\rm max}(T) = 0$, implying that replica
symmetry is restored at the quantum critical point as in the model discussed in reference \onlinecite{Gepasa}. All these conclusions, obtained from the numerical
analysis of
the $p=3$ case, also follow from an approximate analytical solution of
the equations for arbitrary $p\ge3$ \cite{Cugrsa}.

\begin{figure}[7]
\epsfxsize=2.7in
\centerline{\epsffile{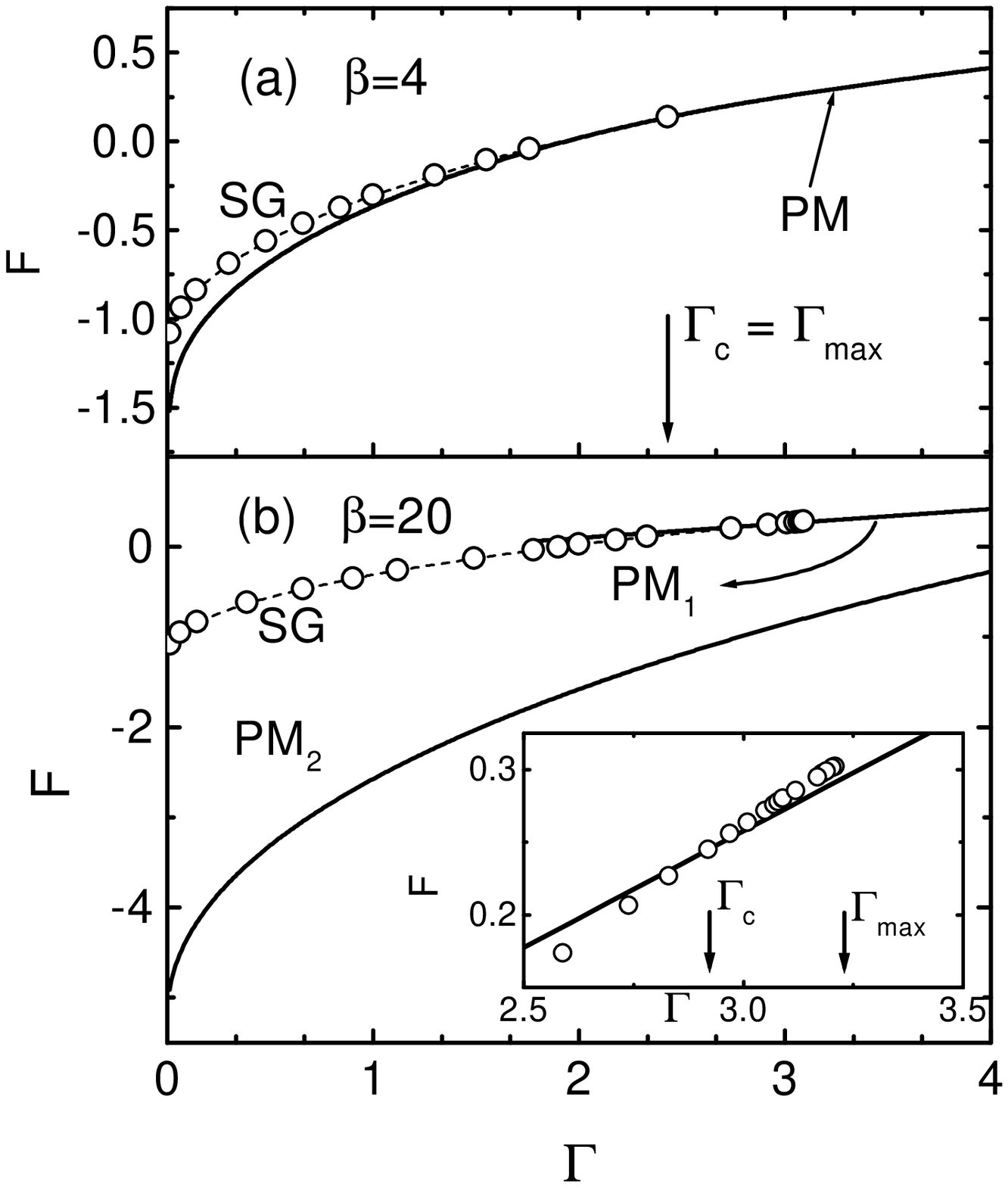}}
\vspace{0.2cm}
\caption{Free-energies of the different PM (solid lines) and SG
(symbols) phases above  (a)
and below (b) $T^{\star}$. The inset in panel (b) shows in detail the
crossing of the free-energies at the critical point for $T < T^{\star}$.}
\label{free-energies}
\end{figure}

PM solutions exist throughout the $T-\Gamma$ plane. The
free-energies of the different states must thus be compared in order to
construct a phase diagram.
Figs.\,\ref{free-energies}(a) and (b) show the $\Gamma$-dependence of
the PM and SG free-energies for the case $p=3$ computed from Eq.~(\ref{eq:free-energy}) for two temperatures,  
above and below $T^{\star}$. Solid lines and symbols
represent the PM and SG solutions, respectively.  The curves end at the point
where the corresponding solution disappears. 
It may be seen that for $T>T^{\star}$  the free-energies of the two
states intersect precisely at $\Gamma_{\rm c}(T)=\Gamma_{\rm max}(T)$: the SG solution
does not extend beyond the transition point and no hysteresis is expected. Below the critical point,
${\rm F}_{\rm SG} >
{\rm F}_{PM}$ meaning that the SG solution
{\it maximizes} the free-energy. This is the usual situation encountered in  replica
theories of classical spin glasses. As in the classical case, $q_{\rm EA}$ is
{\it discontinuous} at the transition. The latter is nevertheless of {\it second}
order because $m=1$ at $\Gamma_{\rm c}$ and, therefore, the effective number of
degrees of freedom participating in the transition $(1-m) q_{\rm EA} \to 0$ at $\Gamma_{\rm c}$. There is no latent heat and
the linear susceptibility is continuous.
\begin{figure}[t]
\epsfxsize=3in
\centerline{\epsffile{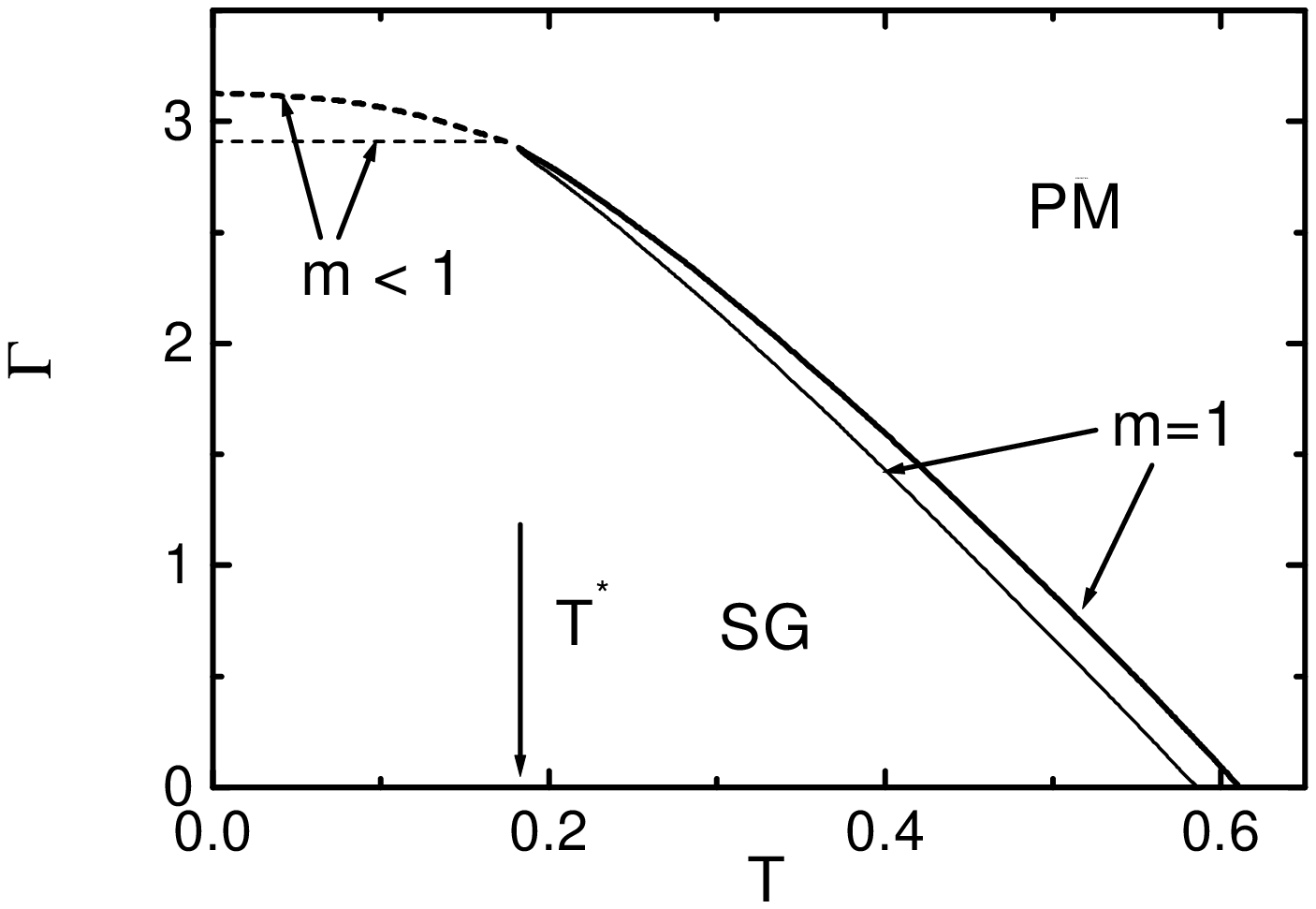}}
\vspace{.2cm}
\caption{Static (thin lines) and dynamic (thick lines) phase diagrams
of the $p$-spin model for $p=3$. Solid and dashed lines represent
second and first 
order transitions, respectively.}
\label{phase}
\end{figure}

The situation is more involved below
$T^{\star}$. 
To start with, one has to choose between the two PM solutions labeled PM$_1$
and PM$_2$ in Fig.\,\ref{free-energies}.
Naively, one would choose the solution with the
lowest free-energy, {\it i.e.}, PM$_2$. However, this solution has
unphysical properties. 
As shown in Fig.~\ref{free-energies}(b), its free-energy {\it never}
 intersects that of the
SG phase. Both the free-energy and the susceptibility diverge as
$T\to 0$. Furthermore,  this solution disappears at a finite value of $\Gamma$ (not shown in
the figure) and cannot thus be reached starting from $\Gamma =
\infty$. On the other hand, it is clear that the ground-state of Hamiltonian (\ref{eq:action}) must have finite susceptibility and energy. We thus conclude that PM$_2$ is a spurious
solution and that PM$_1$ has to be chosen even if its free-energy is {\it higher} \cite{Rogr}. The free-energies of the SG
and PM$_1$ states cross at $\Gamma_{\rm c} <
\Gamma_{\rm max}$ as shown in the inset in Fig.~\ref{free-energies}(b). In the low temperature phase, ${\rm F}_{\rm
SG} < {\rm F}_{\rm PM}$, the opposite of what we found for $T >
T^{\star}$. The SG and PM
solutions extend beyond the point where they cross. There is a region of
phase coexistence and hysteresis effects are thus
expected in the behavior of observables.

\begin{figure}[t]
\epsfxsize=2.7in
\epsfysize=2.7in
\centerline{\epsffile{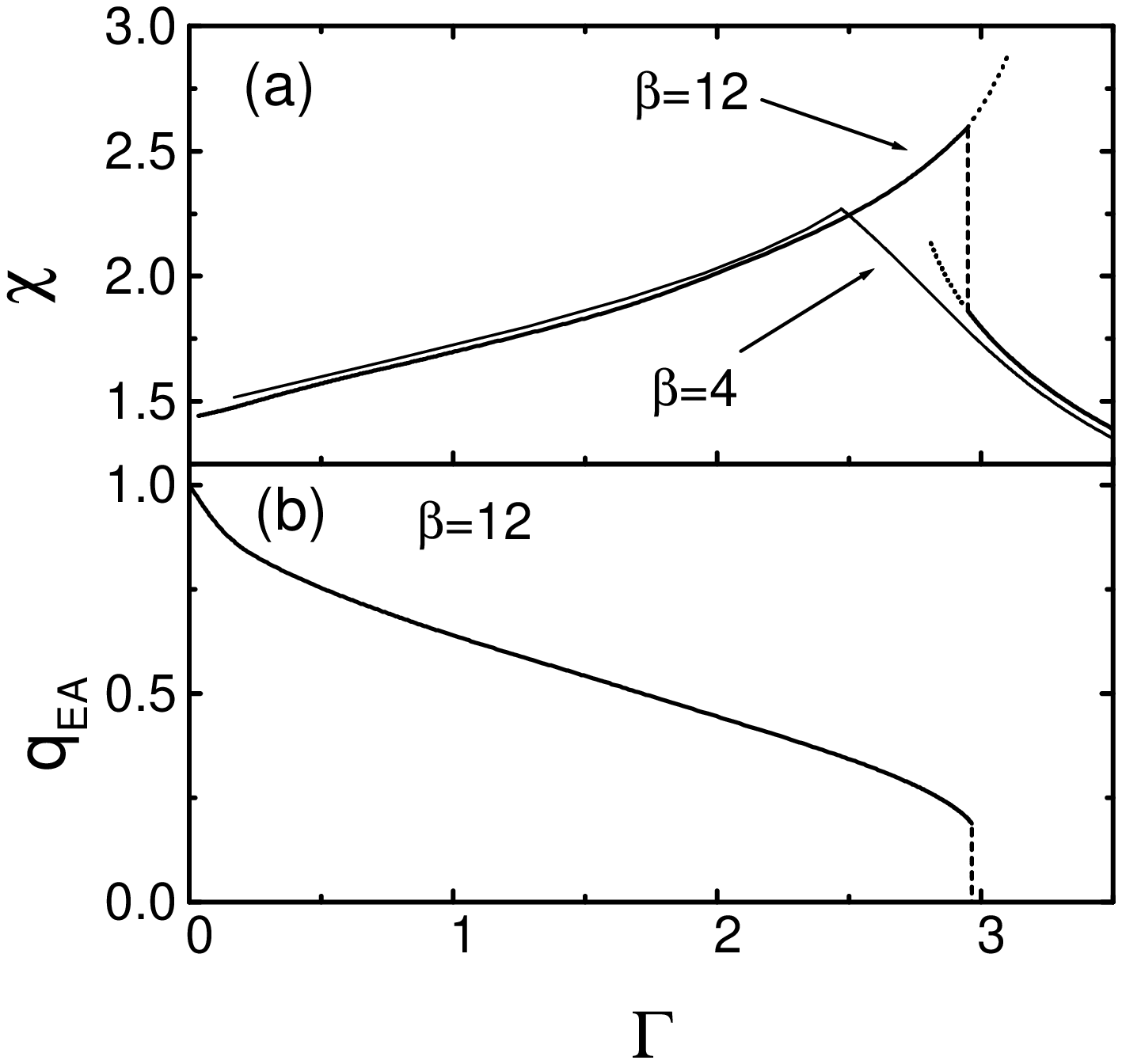}}
\vspace{0.2cm}
\caption{Magnetic susceptibility (a) and Edwards-Anderson order
parameter (b) of the $p$=3 model.}
\label{chi}
\end{figure}

Since now  $q_{\rm EA}$ and $m$ are
discontinuous at $\Gamma_{\rm c}$ the thermodynamic transition is {\it
first} order with latent heat and discontinuous
susceptibility (see below). The phase diagram resulting from this analysis is
represented in Fig.\,\ref{phase} (thin lines). The flat
section is the first-order line. We have computed  the Edwards-Anderson order parameter and the susceptibility, $\chi=\int_0^{\beta} d\tau [q_d(\tau) - (1-m) q_{\rm
EA}]$, as functions of $\Gamma$ for the  $p=3$ model. The results are displayed in Fig.\,\ref{chi}.
The susceptibility has a cusp at $\Gamma_{\rm c}$ for $T > T^{\star}$
and a discontinuity
for $T < T^{\star}$. The dotted lines correspond to the regions of
metastability. Their end points give the amplitude of the ideal
hysteresis cycle.  
The importance
of  quantum
fluctuations may be appreciated from the fact that half way from the
transition the order parameter is already reduced by a factor of two. It can
be shown analytically \cite{Cugrsa} that the spectrum of magnetic
excitations at $T=0$ is
gaped {\it both} in the PM and SG phases for all $\Gamma \ne 0$ (see
below, however). Consequently, the latent heat vanishes exponentially as
$T\to 0$. Since it also vanishes at $T^{\star}$, it must have a
maximum at some intermediate temperature. 

First order quantum transitions
were also found in two other models, the fermionic SK-like spin-glass
model\cite{oppermann} and a $p$-spin model in a transverse
field\cite{Niri}. In contrast, the SG transitions of the Heisenberg EA
model and of the SK model in a transverse field are known to be second
order\cite{daniel}. This is also true in finite dimensions \cite{Fisher,heiko,young}. Early experiments on  LiHo$_x$Y$_{1-x}$F$_4$
gave some indications that the second order SG transition seen above 25 mK, might become 
first order at lower temperatures \cite{aeppli1}. More recently, hysteresis
effects have been observed in this system as a function of the transverse field \cite{aeppli2}, giving further support to this idea. 
The model that we study here captures this phenomenology.

We discuss next the consequences of the use of the marginality condition
\cite{marginality} rather than thermodynamics in the determination of the breaking 
point. In this approach it is  not required that ${\rm F}$ be an
extremum with respect to $m$ but that the SG phase be 
marginally stable. This implies that the ``replicon'' eigenvalue
$\Lambda$ must vanish throughout the low-temperature phase.  
The calculation of $\Lambda$ \cite{Cugrsa} is analogous to the classical
one \cite{Crso}. The result is

\begin{eqnarray}
\label{lambda}
\Lambda &=& \left[
\frac{
\tilde q_d(0)+\beta q_{\rm EA}(m-1)
}{
\tilde q_d^2(0)-\beta^2 q_{\rm EA}^2 (m-1)+\tilde q_d(0) \beta q_{\rm
EA}(m-2)}\right]^2 \beta^2
\nonumber \\
& & 
-\frac{\beta^2}{2} p (p-1) q_{\rm EA}^{p-2}
\; .
\end{eqnarray}
The value of $m$ follows from the equation $\Lambda=0$, that combined with the 
equation $\delta {\rm F}/\delta q_{\rm EA} = 0$, yields  
\begin{equation}
\label{eq:thetam}
m = T (p-2) \sqrt{2/(p(p-1))} \; q_{\rm EA}^{-p/2}.
\end{equation}
Notice that this expression is equivalent to Eq.\,(\ref{eq:theta}) 
with the substitution $x_p \to (p-2)$. Interestingly enough, Eq.\,(\ref{eq:thetam}) is identical to the equation found for the fluctuation-dissipation theorem (FDT) violation parameter, $X$, 
in the real-time dynamical
calculation\,\cite{Culo}. Moreover, the static and dynamical equations for $q_{\rm EA}$ are also identical which implies that $m=X$.  The coincidence between the
values of $X$ and $m$ for the  marginal SG state has been 
noticed several times for classical models. This is the first explicit
evidence 
of its validity in a quantum problem. $X$ is related to the effective
temperature \cite{CuKuPe} of the system, $T_{\rm eff}= X^{-1} T$, where $T$ is the
temperature of the thermal bath it is in contact with. Values of
$X\ne 1$ signal the presence of a non-stationary dynamics and of FDT violations. The fact that $\beta X=\beta m \to {\rm const}$ when $T\to 0$ shows that a non-trivial $T_{\rm eff}$ is generated even when the temperature of the bath vanishes. We have also 
shown analytically\,\cite{Cugrsa} that the internal energy, computed from
$U=\partial(\beta {\rm F})/\partial \beta$ at {\it constant} $m$,
 coincides with the long-time limit of the
energy per spin as obtained from dynamics \cite{Culo}. The dynamic
transition line $\Gamma_{\rm d}(T)$ may be thus identified as the boundary of the region in
the $T-\Gamma$ plane where the marginally stable SG exists.
Below this line, the dynamics of the system becomes 
non-stationary and FDT violations set in. 
 The dynamic phase diagram for $p=3$ is
shown in Fig.\,\ref{phase} (thick lines). 
As in the equilibrium case, $m$ is discontinuous across the dashed line. $\Gamma_{\rm d}$ lies always 
above $\Gamma_{\rm c}$ suggesting that the equilibrium state can
never be reached dynamically starting from an initial state in the PM
phase. 
The two lines are extremely close to each
other for $T \sim T^{\star}$. Within the accuracy of our calculations
we cannot assert whether they precisely touch at $T^{\star}$, an intriguing possibility. 
In the region $T < T^{\star}$, $m$ varies continuously along
$\Gamma_{\rm d}(T)$ and vanishes at the quantum
critical point. This has a consequence of potential interest for
experiment: FDT violations are predicted to appear suddenly
rather than gradually as $\Gamma_{\rm d}$ is crossed coming
from the high $\Gamma$ region for $T < T^{\star}$.  The stationary part of the
time-dependent susceptibility in the SG phase can be calculated by
analytic continuation of $\tilde{q}'_d(\omega_k)$. It may be shown that the excitation spectrum of the marginal SG
state is gapless \cite{Cugrsa}. Furthermore, $\chi''(\omega)$ may be
calculated exactly for $T, \omega \to 0$. The result is
\begin{equation}
\label{eq:spectrum}
\lim_{\omega\to 0}\frac{\chi''(\omega)}{\omega} = 
\frac{1}{\Gamma} \left[\frac{2\,q_{\rm EA}^{(2-p)}}{p (p-1)}\right]^{3/4}.
\end{equation}
A linear excitation
spectrum has also been found in the case of the Heisenberg spin glass
model\,\cite{Gepasa}. However, a gapless spectrum is not a  
consequence of Goldstone's theorem here as our model does not have any continuous symmetry.
A more extended discussion of our results will be presented in a forthcoming paper \cite{Cugrsa}.

One of us (CAdSS) acknowledges financial support from 
the Portuguese Research Council, FCT, under grant BPD/16303/98. LFC and DRG thank the ECOS-Sud program for a travel grant. 
We thank T. F. Rosenbaum for making available to us the results of
Ref. \onlinecite{aeppli2} prior to publication. We also thank
G. Biroli, J. Kurchan, G. Lozano and M. Rozenberg for useful 
discussions.

\end{document}